\documentclass[runningheads]{llncs}
\usepackage{graphicx}
\usepackage{units}
\usepackage{todonotes}
\usepackage{booktabs}
\usepackage{manfnt}
\usepackage{amsmath,amssymb}
\usepackage{mathtools}
\usepackage{blindtext}
\usepackage{xcolor}
\usepackage{bm}

\newcommand{\prn}[1]{\left(#1\right)}
\newcommand{\ud}[1]{\, \mathrm{d}#1}
\newcommand{\R}{\mathbb{R}}
\newcommand{\StaRMAP}{{StaRMAP}} 
\newcommand{\NAStJA}{{NAStJA}}
\newcommand{\MATLAB}{{MATLAB}}

\begin{document}
\title{Massively Parallel Stencil Strategies for Radiation Transport Moment Model Simulations}
\titlerunning{Massively Parallel Stencil Strategies for Radiation Transport}
%
\author{Marco Berghoff\inst{1}\orcidID{0000-0003-4343-2228} \and
Martin Frank\inst{1}\orcidID{0000-0001-8562-6982} \and
Benjamin Seibold\inst{2}\orcidID{0000-0003-2879-6402}}
\authorrunning{M. Berghoff et al.}
%
\institute{Steinbuch Centre for Computing, Karlsruhe Institute of Technology, Karlsruhe, Germany
\email{\{marco.berghoff, martin.frank\}@kit.edu}\\
\and
Department of Mathematics, Temple University, Philadelphia PA 19122, USA \\
\email{seibold@temple.edu}}
\maketitle              
%
\begin{abstract}
The radiation transport equation is a mesoscopic equation in high dimensional phase space.
Moment methods approximate it via a system of partial differential equations in traditional space-time.
One challenge is the high computational intensity due to large vector sizes (1\,600 components for P39) in each spatial grid point.
In this work, we extend the calculable domain size in 3D simulations considerably, by implementing the \StaRMAP{} methodology within the massively parallel HPC framework \NAStJA{}, which is designed to use current supercomputers efficiently.
We apply several optimization techniques, including a new memory layout and explicit SIMD vectorization.
We showcase a simulation with 200 billion degrees of freedom, and argue how the implementations can be extended and used in many scientific domains.

\end{abstract}

\section{Introduction}
The accurate computation of radiation transport is a key ingredient in many application problems, including astrophysics~\cite{Mihalas,ZeldovichRaizer,Pomraning}, nuclear engineering~\cite{Murray,Davison,CaseZweifel}, climate science~\cite{DavisMarshak}, nuclear medicine~\cite{Lar97}, and engineering~\cite{Modest}.
A key challenge for solving the (energy-independent) radiation transport equation (RTE) \eqref{eq:RTE} is that it is a mesoscopic equation in a phase space of dimension higher than the physical space coordinates.
Moment methods provide a way to approximate the RTE via a system of macroscopic partial differential equations (PDEs) defined in traditional space-time. Here we consider the $P_N$ method~\cite{BrunnerHolloway2005}, which is based on an expansion of the solution of \eqref{eq:RTE} in Spherical Harmonics. It can be interpreted as a moment method or, equivalently, as a spectral semi-discretization in the angular variable. Advantages of the $P_N$ method over angular discretizations by collocation (discrete ordinates, $S_N$)~\cite{MorelWareingLowrieParsons2003} is that it preserves rotational invariance. A drawback, particular in comparison to nonlinear moment methods~\cite{Kershaw1976,Levermore1984,Su2001,AnilePennisiSammartino1991,MullerRuggeri1993,TurpaultFrankDubrocaKlar2004}, are spurious oscillations (``wave effects'') due to Gibbs phenomena.
To keep these at bay, it is crucial that the $P_N$ method be implemented in a flexible fashion that preserves efficiency and scalability and allows large values of $N$.

Studies and applications of the $P_N$ methods include~\cite{BrunnerHolloway2005,McClarrenHollowayBrunner2008,McClarrenEvansDensmore2008,Olson2009}.
An important tool for benchmarking and research on linear moment methods is the \StaRMAP{} project~\cite{SeiboldFrank2014}, developed by two authors of this contribution.
Based on a staggered grid stencil approach (see \S\ref{subsec:numerical_methodology}), the \StaRMAP{} approach is implemented as an efficiently vectorized open-source \MATLAB{} code~\cite{StaRMAP}.
The software's straightforward usability and flexibility have made it a popular research tool, used in particular in numerous dissertations, and it has been extended to other moment models (filtered~\cite{HauckMcClarren2010} and simplified~\cite{OlbrantLarsenFrankSeibold2013}), and applied in radiotherapy simulations~\cite{FrankKuepperSeibold2014,KuepperDiss2016}.
For 2D problems, the vectorized \MATLAB{} implementation allows for serial or shared memory (\MATLAB{}'s automatic usage of multiple cores) parallel execution speeds that are on par with comparable implementations of the methodology in C++. The purpose of this paper is to demonstrate that the same \StaRMAP{} methodology also extends to large-scale, massively parallel computations and yields excellent scalability properties.

While $S_N$ solvers for radiation transport are important production codes and major drivers for method development on supercomputers (one example is DENOVO~\cite{denovo}, which is one of the most time-consuming codes that run in production mode on the Oak Ridge Leadership Computing Facility~\cite{Messer2018}), we are aware of only one work~\cite{GarrettHauckHill2015} that considers massively parallel implementations for moment models.

The enabler to transfer \StaRMAP{} to current high-performance computing (HPC) systems is the open-source \NAStJA{} framework~\cite{NAStJA,BerghoffKondovHoetzer2018}, co-developed by one author of this contribution.
\NAStJA{} is a massively parallel framework for stencil-based algorithms on block-structured grids.
The framework has been shown to efficiently scale up to more than ten thousand threads~\cite{BerghoffKondovHoetzer2018} and run simulations in several areas, using the phase-field method for water droplets~\cite{berghoff2018non}, the phase-field crystal model for crystal--melt interfaces~\cite{guerdane2018crystal} and cellular Potts models for tissue growth and cancer simulations~\cite{berghoff2019massively} with millions of grid points.

\section{Model}
The radiation transport equation (RTE)~\cite{CaseZweifel}
\begin{equation}
\label{eq:RTE}
\begin{multlined}
\partial_t \psi(t,x,\Omega)
+ \Omega\cdot\nabla_x\psi(t,x,\Omega)
+ \Sigma_t(t,x)\psi(t,x,\Omega)\\
= \int_{S^2} \Sigma_s(t,x,\Omega\cdot\Omega') \psi(t,x,\Omega')\ud{\Omega'} + q(t,x,\Omega),
\end{multlined}
\end{equation}
equipped with initial data $\psi(0,x,\Omega)$ and suitable boundary conditions, describes the evolution of the density $\psi$ of particles undergoing scattering and absorption in a medium (units are chosen so that the speed of light $c = 1$).
The phase space consists of time $t>0$, position $x\in\R^3$, and flight direction $\Omega\in S^2$.
The medium is characterized by the cross-section $\Sigma_t$ (see below) and scattering kernel $\Sigma_s$.
Equation \eqref{eq:RTE} stands representative for more general radiation problems, including electron and ion radiation~\cite{FrankHertySchaefer2008} and energy-dependence~\cite{LarsenMiftenFraassBruinvis1997}.

Moment methods approximate \eqref{eq:RTE} by a system of macroscopic equations.
In 1D slab geometry, expand the $\Omega$-dependence of $\psi$ in a Fourier series, $\psi(t,x,\mu) = \sum_{\ell=0}^\infty \psi_\ell(t,x) \tfrac{2\ell+1}{2} P_\ell(\mu)$, where $\mu$ is the cosine of the angle between $\Omega$ and $x$-axis, and $P_\ell$ are the Legendre polynomials.
Testing \eqref{eq:RTE} with $P_\ell$ and integrating yields equations for the Fourier coefficients $\psi_\ell = \int_{-1}^1 \psi P_\ell \ud{\mu}$ as
\begin{equation}
\label{eq:moment_evolution}
\partial_t \psi_\ell+\partial_x\int_{-1}^1 \mu P_\ell \psi \ud{\mu} + \Sigma_{t\ell} \psi_\ell =  q_\ell
\quad\quad\text{for~}\ell=0,1,\dots\;,
\end{equation}
where $\Sigma_{t\ell} = \Sigma_t - \Sigma_{s\ell} = \Sigma_a + \Sigma_{s0} - \Sigma_{s\ell}$ and $\Sigma_{s\ell} = 2\pi \int_{-1}^1 P_\ell(\mu) \Sigma_s(\mu) \ud{\mu}$.
Using the three-term recursion for Legendre polynomials, relation \eqref{eq:moment_evolution} becomes
\begin{equation*}
\partial_t \psi_\ell+\partial_x \prn{ \tfrac{\ell+1}{2\ell+1}\psi_{\ell+1}
+ \tfrac{\ell}{2\ell+1}\psi_{\ell-1} } + \Sigma_{t\ell} \psi_\ell = q_\ell.
\end{equation*}
These equations can be assembled into an infinite system $\partial_t\vec{u} + M\cdot\partial_x\vec{u} +C\cdot\vec{u} = \vec{q}$, where $\vec{u} = (\psi_0,\psi_1,\dots)^T$ is the vector of moments, $M$ is a tri-diagonal matrix with zero diagonal, and $C = \text{diag}(\Sigma_{t0},\Sigma_{t1},\dots)$ is diagonal.
The slab-geometry $P_N$ equations are now obtained by omitting the dependence of $\psi_N$ on $\psi_{N+1}$ (alternative interpretations in~\cite{LarsenMorelMcGhee1996,SeiboldFrank2009,FrankSeibold2011}).

In 2D and 3D, there are multiple equivalent ways to define the $P_N$ equations (cf.~\cite{BrunnerHolloway2005,SeiboldFrank2014}).
\StaRMAP{} is based on the symmetric construction using the moments $\psi_\ell^m(t,x) = \int_{S^2} \overline{Y_\ell^m(\Omega)} \psi(t,x,\Omega) \ud{\Omega}$, with the complex spherical harmonics $Y_\ell^m(\mu,\varphi) = (-1)^m \sqrt{\tfrac{2\ell+1}{4\pi}\tfrac{(\ell-m)!}{(\ell+m)!}}\, e^{im\varphi}P_\ell^m(\mu)$, where $\ell\geq 0$ is the moment order, and $-\ell\leq m\leq \ell$ the tensor components.
Appropriate substitutions~\cite{SeiboldFrank2014} lead to real-valued $P_N$ equations.
In 3D the moment system becomes
\begin{equation}
\label{eq:hyperbolic_balance_law}
\partial_t\vec{u} + M_x\cdot\partial_x\vec{u} + M_y\cdot\partial_y\vec{u}
+ M_z\cdot\partial_z\vec{u} + C\cdot\vec{u} = \vec{q}\;,
\end{equation}
where the symmetric system matrices $M_x$, $M_y$, $M_z$ are sparse and possess a very special pattern of nonzero entries (see~\cite{SeiboldFrank2014,StaRMAP}).
That coupling structure between unknowns (same in 2D and 1D) enables elegant and effective staggered grid discretizations upon which \StaRMAP{} is based.

\subsection{Numerical Methodology}
\label{subsec:numerical_methodology}
We consider the moment system \eqref{eq:hyperbolic_balance_law} in a rectangular computational domain $(0,L_x)\times (0,L_y)\times (0,L_z)$ with periodic boundary conditions (see below).
The domain is divided into $n_x \times n_y\times n_z$ rectangular equi-sized cells of size $\Delta x \times \Delta y \times \Delta z$.
The center points of these cells lie on the base grid
\begin{equation*}
G_{111} = \left\{\left(\left(i\!-\!\tfrac{1}{2}\right)\Delta x, \left(j\!-\!\tfrac{1}{2}\right)\Delta y, \left(k\!-\!\tfrac{1}{2}\right)\Delta z\right) \mid (1,1,1)\le (i,j,k)\le (n_x,n_y,n_z) \right\}.
\end{equation*}
The first component of $\vec{u}$ (the zeroth moment, which is the physically meaningful radiative intensity) is always placed on $G_{111}$.
The other components of $\vec{u}$ are then placed on the 7 other staggered grids
$G_{211} = \left\{\left(i\Delta x, \left(j\!-\!1/2\right)\Delta y, \left(k\!-\!1/2\right)\Delta z\right)\right\}$,
$G_{121} = \left\{\left(\left(i\!-\!1/2\right)\Delta x, j\Delta y, \left(k\!-\!1/2\right)\Delta z\right)\right\}$, \dots,
$G_{222} = \left\{\left(i\Delta x, j\Delta y, k\Delta z\right)\right\}$, following the fundamental principle that an $x$-derivative of a component in \eqref{eq:hyperbolic_balance_law} that lives on a $(1,\bullet,\bullet)$ grid updates a component that lives on the corresponding $(2,\bullet,\bullet)$ grid.
Likewise, $x$-derivatives of components on $(2,\bullet,\bullet)$ grids update information on the $(1,\bullet,\bullet)$ grids; and analogously for $y$- and $z$-derivative.
A key result, proved in~\cite{SeiboldFrank2014}, is that this placement is, in fact, always possible.

Due to this construction, all spatial derivatives can be approximated via simple second-order half-grid centered finite difference stencils: two $x$-adjacent values, for instance living on the $(1,1,1)$ grid, generate the approximation
\begin{equation*}
\partial_x u (i\Delta x, (j\!-\!\tfrac{1}{2})\Delta y, (k\!-\!\tfrac{1}{2})\Delta z)
= \frac{
u_{(i\!+\!\frac{1}{2}, j\!-\!\frac{1}{2}, k\!-\!\frac{1}{2})}
-u_{(i\!-\!\frac{1}{2}, j\!-\!\frac{1}{2}, k\!-\!\frac{1}{2})}
}{\Delta x}
+ O(\Delta x^2)
\end{equation*}
on the $(2,1,1)$ grid. We now call the $G_{111}$, $G_{221}$, $G_{122}$, and $G_{212}$ grids ``even'', and the $G_{211}$, $G_{121}$, $G_{112}$, and $G_{222}$ grids ``odd''.

The time-stepping of \eqref{eq:hyperbolic_balance_law} is conducted via bootstrapping between the even and the odd grid variables.
This is efficiently possible because of the approximate spatial derivatives of the even/odd grids update \emph{only} the components that live on the odd/even grids.
Those derivative components on the dual grids are considered ``frozen'' during a time-update of the other variables, leading to the decoupled update ODEs
\begin{equation}
\label{eq:hyperbolic_balance_law_decoupled}
\begin{cases}
\partial_t\vec{u}^\text{e}+C^\text{e}\cdot\vec{u}^\text{e}
= \vec{q}^\text{e} - (M_x^\text{eo}\cdot D_x + M_y^\text{eo}\cdot D_y + M_z^\text{eo}\cdot D_z)\vec{u}^\text{o} \\
\partial_t\vec{u}^\text{o}+C^\text{o}\cdot\vec{u}^\text{o}
= \vec{q}^\text{o} - (M_x^\text{oe}\cdot D_x + M_y^\text{oe}\cdot D_y + M_z^\text{oe}\cdot D_z)\vec{u}^\text{e}
\end{cases}
\end{equation}
for the vector of even moments $\vec{u}^\text{e}$ and the vector of odd moments $\vec{u}^\text{o}$.
In \eqref{eq:hyperbolic_balance_law_decoupled}, the right-hand sides are constant in time (due to the freezing of the dual variables, as well as the source $\vec{q}$).
Moreover, because $C^\text{e}$ and $C^\text{o}$ are diagonal, the equations in \eqref{eq:hyperbolic_balance_law_decoupled} decouple further into scalar ODEs of the form
\begin{equation*}
\partial_t u_k(\vec{x},t) + \bar{c}_k(\vec{x})u_k(\vec{x},t) = \bar{r}_k(\vec{x}),
\end{equation*}
whose exact solution is
\begin{equation}
\label{eq:sub_step_solution}
u_k(\vec{x},t+\Delta t)
= u_k(\vec{x},t) + \Delta t\prn{\bar{r}_k(\vec{x}) - \bar{c}_k(\vec{x}) u_k(\vec{x},t)}
E(-\bar{c}_k(\vec{x})\Delta t).
\end{equation}
Here $\vec{x} = (x,y,z)$ is the spatial coordinate, and $E(c) = (\exp(c)-1)/c$ (see~\cite{SeiboldFrank2014} for a robust implementation of this function).
To achieve second order in time, one full time-step (from $t$ to $t+\Delta t$) is now conducted via a Strang splitting
\begin{equation}
\label{eq:strang_splitting}
\vec{u}(\vec{x},t+\Delta t) =
S_{\frac{1}{2}\!\Delta t}^\text{o}\circ
S_{\Delta t}^\text{e}\circ
S_{\frac{1}{2}\!\Delta t}^\text{o}
\vec{u}(\vec{x},t),
\end{equation}
where $S_{\frac{1}{2}\Delta t}^\text{o}$ is the half-step update operator for the odd variables, and $S_\Delta t^\text{e}$ the full-step update operator for the even variables, both defined via \eqref{eq:sub_step_solution}.

The convergence of this method, given that $\Delta t<\min\{\Delta x, \Delta y, \Delta z\}/3$, has been proven in~\cite{SeiboldFrank2014}.
Stability is generally given even for larger time-steps if scattering is present.



\section{Implementation}
In the following section, we present our implementation of the \StaRMAP{} model and applied optimizations that are required to run on current HPC systems efficiently.
\StaRMAP{} v1.0~\cite{StaRMAP2020} is published under the Mozilla Public License 2.0 and the source-code is available at \url{https://gitlab.com/nastja/starmap}.

\subsection{The \NAStJA{} Framework}
The \StaRMAP{} methodology described above was implemented using the open-source \NAStJA{} framework\footnote{The MPL-2.0 source-code is available at \url{https://gitlab.com/nastja/nastja}.}.
The framework was initially developed to explore non-collective communication strategies for simulations with a large number of MPI ranks, as will be used in exascale computing.
It was developed in such a way that many multi-physics applications based on stencil algorithms can be efficiently implemented in a parallel way.
The entire domain is build of blocks in a block-structured grid. These blocks are distributed over the MPI ranks.
Inside each block, regular grids are allocated for the data fields.
The blocks are extended with halo layers that hold a copy of the data from the neighboring blocks.
This concept is flexible, so it can adaptively create blocks where the computing area moves.
The regular structure within the blocks allows high-efficiency compute kernels, called sweeps.
Every process holds information only about local and adjacent blocks. The framework is entirely written in modern C++ and makes use of template metaprogramming to achieve excellent performance without losing flexibility and usability.
Sweeps and other actions are registered and executed by the processes in each time-step for their blocks so that functionality can be easily extended.
Besides, sweeps can be replaced by optimized sweeps, making it easy to compare the optimized version with the initial one.

\subsection{Optimizations}\label{sec:opti}
Starting with the 3D version of the \MATLAB{} code of \StaRMAP{}, the goal of this work was to develop a highly optimized and highly parallel code for future real-time simulations of radiation transports.

\emph{Basic implementation.}
The first step was to port the \MATLAB{} code to C++ into the \NAStJA{} framework.
Here we decide to use spatial coordinates ($x,y,z$) as the underlying memory layout.
At each coordinate, the vector of moments $\vec{u}$ is stored.
The sub-grids $G_{111}$ to $G_{222}$ are only considered during the calculation and are not saved separately.
This means that the grid points on $G_{111}$ and all staggered grid points are stored at the non-staggered $(x, y, z)$-coordinates.
Thus it can be achieved that data that are needed for the update is close to each other in the memory.
As for Equation~\eqref{eq:hyperbolic_balance_law_decoupled} described, all even components are used to calculate the odd components and vice versa.
This layout also allows the usage of a relatively small stencil.
The D3C7 stencil, which reads for three dimensions, the central data point and the first six direct neighbors, is sufficient.

For parallelization, we use \NAStJA{}'s block distribution and halo exchange mechanisms.
The halo is one layer that holds a copy of the $\vec{u}$ vectors from the neighboring blocks.
Since a D3C7 stencil is used, it is sufficient to exchange the six first neighboring sides.
\begin{figure}[tb!]
  \centering
  \includegraphics[width=0.48\textwidth]{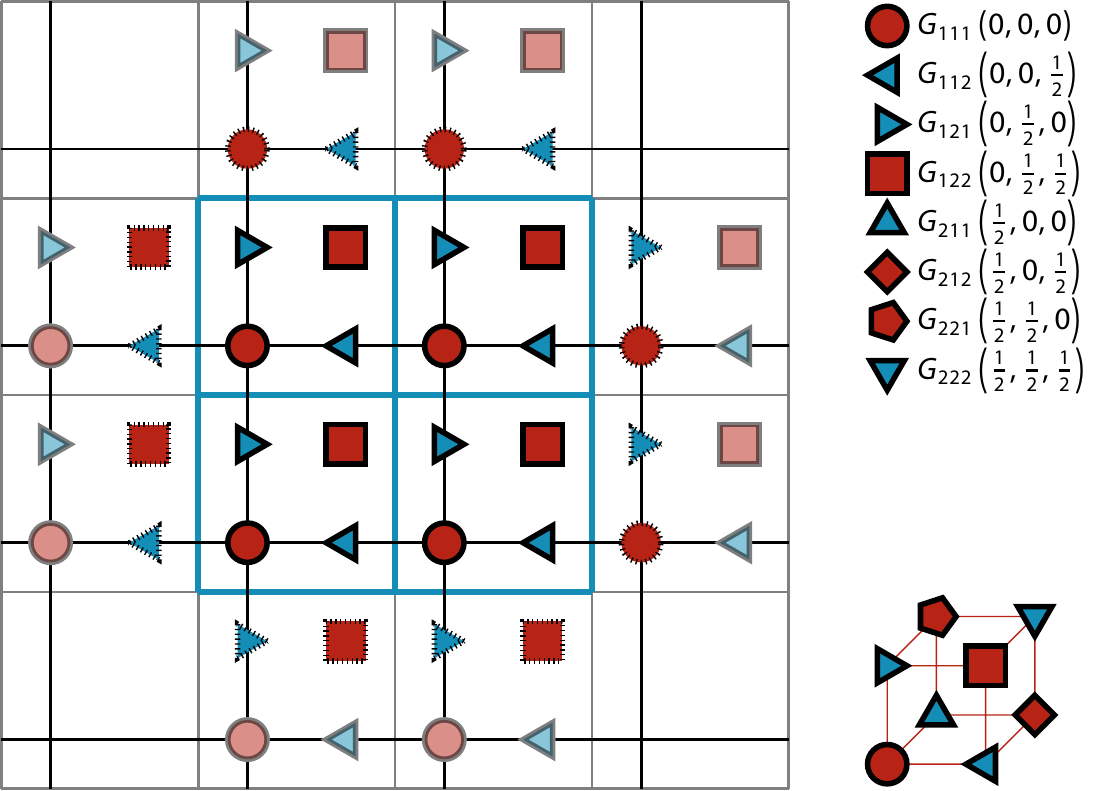}\hfill
  \includegraphics[width=0.42\textwidth]{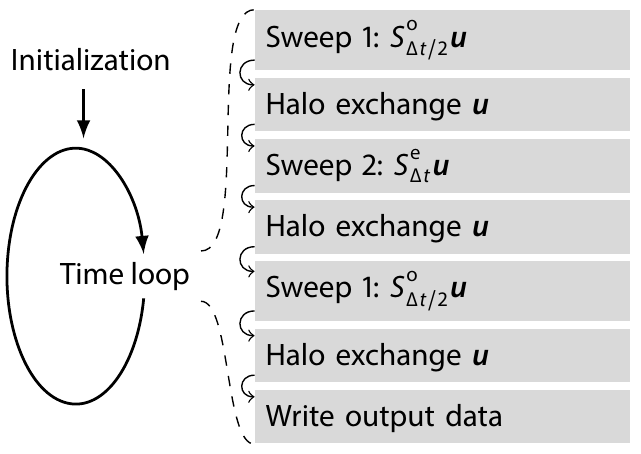}
  \caption{Left: Staggered grids for the first $z$-layer.
  The odd coordinates are blue triangles and the even coordinates are marked by red shapes.
  The \NAStJA{}-cells are blue squares.
  The base (111) grid is denoted by the black lines.
  The grid points with dotted border are the halo layer or the periodic boundary copy, the light grid points are not used.
  Center: 3D \NAStJA{}-cell with the base grid point $G_{111}$ (red circle) and the seven staggered grid points.
  Right: Action and sweep setup in \NAStJA.}
  \label{fig:grid}
\end{figure}
Fig.~\ref{fig:grid} left shows the grid points in \NAStJA{}'s cells and the halo layer.
For the implemented periodic boundary condition, we use this halo exchange to copy \NAStJA{}-cells from one side to the opposite side, even if only half of the moments are needed to calculate the central differences.

For the calculation of the four substeps in Equation~\eqref{eq:strang_splitting}, two different sweeps are implemented, each sweep swipes over the spatial domain in $z,y,x$ order.
The updates of each $\vec{u}$ component for each cell is calculated as followed.
Beginning with the first substep, sweep $S^\text{o}$ calculates $d_x\vec{u}, d_y\vec{u}, d_z\vec{u}$ of the even components as central differences, laying on the odd components.
Then, the update of the odd components using this currently calculated $d_x\vec{u}, d_y\vec{u}, d_z\vec{u}$ is calculated.
After the halo layer exchange, sweep $S^\text{e}$ calculates the second substep.
Therefore, first, the $d_x\vec{u}, d_y\vec{u}, d_z\vec{u}$ of the odd components are calculated, followed by the update of the even components.
A second halo layer exchange proceeds before the sweep $S^\text{o}$ is called again to complete with the third substep.
The time-step is finalized by a third halo layer exchange and an optional output.
Fig.~\ref{fig:grid} right shows the whole sweep setup of one time-step in the \NAStJA{} framework.

The time-independent parameter values as $\vec{q}$, $\bar{c}_k(\vec{x})$, and $E(-\bar{c}_k(\vec{x})\Delta t/2)$  are stored in an extra field on the non-staggered coordinates.
Here, $\bar{c}_k(\vec{x})$ for $k\geq1$ are identical.
Their values on the staggered grid positions are interpolated.

\emph{Reorder components.}
For optimization purposes, the calculation sweeps can easily exchange in \NAStJA.
Two new calculation sweeps are added for each of the following optimization steps.
The computational instructions for the finite differences of the components on one sub-grid are the same, as well as the interpolated parameter values.
Components of the vector $\vec{u}$ are reordered, in that way that components of individual sub-grids are stored sequentially in memory.
First, the even then, the odd sub-grid components follow, namely $G_{111}$, $G_{221}$, $G_{212}$, $G_{122}$, $G_{211}$, $G_{121}$, $G_{112}$, and $G_{222}$.

\emph{Unroll multiplications.}
The calculation of $w = M_x \cdot d_x \vec{u} + M_y \cdot d_y \vec{u} + M_z \cdot d_z \vec{u}$ is optimized by manually unroll and skipping multiplication.
The Matrices $M_x$ and $M_y$ have in each row one to four non-zero entries while the Matrix $M_z$ has zero to two non-zero entries.
Only these non-zero multiplication have to sum up to $w$.
The first if-conditions for the non-zero entries in $M_x$ and $M_y$ is always true so that it can be skipped.
A manual loop-unroll with ten multiplications and eight if-conditions is used.

\emph{SIMD intrinsics.}
The automatic vectorization by the compilers results in worse run times.
So we decide to manually instruct the code with intrinsics using the Advanced Vector Extensions 2 (AVX2), as supported by the test systems.
Therefore, we reinterpret the four-dimensional data field $(z,y,x,u)$ as a fifth-dimensional data field $(z,y,X,u,x')$, where $x'$ holds the four $x$ values that fit into the AVX vector register, and $X$ is the $x$-dimension shrink by factor $4$.
Currently, we only support multiples of $4$ for the $x$-dimension.
The changed calculation sweeps allow calculating four neighbored values at once.
The fact that the studied number of moments are multiples of $4$ ensures that all the memory access are aligned.
With this data layout, we keep the data very local and can still benefit from the vectorization.

\section{HPC System}
To perform the scaling test, we use a single node (kasper) and the high-per\-for\-mance computing systems ForHLR II, located at Karlsruhe Institute of Technology (fh2).
The single node has two quad-core Intel Xeon processors E5-2623 v3 with Haswell architecture running at a base frequency of $\unit[3]{GHz}$ ($\unit[2.7]{GHz}$ AVX), and have $4 \times \unit[256]{KB}$ of level 2 cache, and $\unit[10]{MB}$ of shared level 3 cache.
The node has \unit[54]{GB} main memory.

The ForHLR II has 1152 20-way Intel Xeon compute nodes~\cite{fh2}.
Each of these nodes contains two deca-core Intel Xeon processors E5-2660 v3 with Haswell architecture running at a base frequency of $\unit[2.6]{GHz}$ ($\unit[2.2]{GHz}$ AVX), and have $10 \times \unit[256]{KB}$ of level 2 cache, and $\unit[25]{MB}$ of shared level 3 cache.
Each node has $\unit[64]{GB}$ main memory, and an FDR adapter to connect to the InfiniBand 4X EDR interconnect.
In total, 256 nodes can be used, which are connected by a quasi fat-tree topology, with a bandwidth ratio of 10:11 between the switches and leaf switches.
The leaf switches connect 23 nodes. The implementation of Open MPI in version 3.1 is used.


\section{Results and Discussion}
In this section, we present and discuss single core performance results as well as scaling experiments run on a high-performance computing system.
The presented performance results are measured in $\unit{MLCUP/s}$, which stands for ``million lattice cell component updates per second''.
This unit takes into account that the amount of data depends on the number of lattice cells and the number of moments.

\subsection{Performance Results}

\subsubsection{Single Core Performance}
The starting point of our HPC implementation was a serial \MATLAB{} code.
A primary design goal of \StaRMAP{} is to provide a general-purpose code with several different functions.
In this application, we focus on specific cases, but let the number of moments be a parameter.
A simple re-implementation in the \NAStJA{} framework yields the same speed as the \MATLAB{} code but has the potential to run in parallel and thus exceed the \MATLAB{} implementation.

\begin{figure}[tb!]
  \centering
  \includegraphics[width=0.9\textwidth]{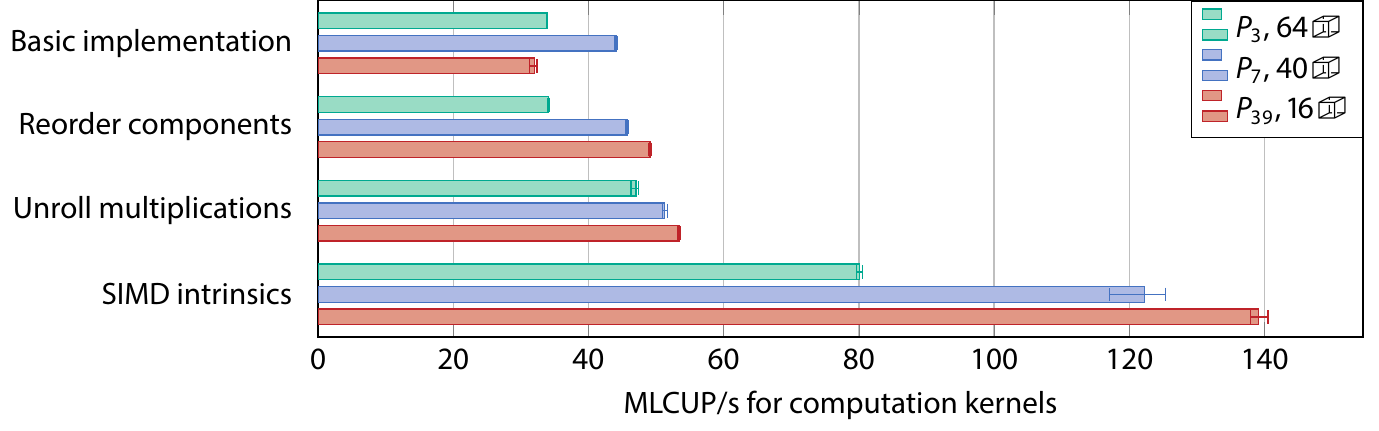}
  \caption{Performance of the various optimization variants of the calculation sweeps running on a single core.
  The block size (\mancube) was chosen so that the number of the total components is approximately equal for all number of moments $M_\bullet$.
  The marks denote the average of three runs.
  The error bars indicate the minimum and maximum.}
  \label{fig:variants}
\end{figure}
Fig.~\ref{fig:variants} shows the performance of the optimization describes in \S\ref{sec:opti}.
The measurements based on the total calculation sweep time per time-step, i.e., two sweep $S^\text{o}$ + sweep $S^\text{e}$.
In all the following simulations, we use cubic blocks, such that a block size of 40 refers to a cubic block with an edge length of 40 lattice cells without the halo.
In legends, we write 40 \mancube.
The speedup from the basic implementation to the reorder components version is small for $P_3$ and $P_7$ but significant for $P_{39}$ ($+\unit[54]{\%}$).
The number of components on each subgrid is small for the first both but large for $P_{39}$, so the overhead of the loops over all components becomes negligible.
Unrolling brings an additional speedup of $\unit[38]{\%}$ for $P_3$, $\unit[14]{\%}$ for $P_7$, and $\unit[9]{\%}$ for $P_{39}$.
Vectorization has the smallest effect for $P_3$ ($+\unit[70]{\%}$).
For $P_7$ we gain $+\unit[138]{\%}$ and $+\unit[160]{\%}$ for $P_{39}$

The combination of all optimizations results in a total speedup of factor $2.36$, $2.77$, $4.35$ for $P_3$, $P_7$, $P_{39}$, respectively.
This optimization enables us to simulate sufficiently large domains in a reasonable time to obtain physically meaningful results.
Note, these results run with a single thread, so the full L3 cache is used.
Since the relative speedup does not indicate the utilization of a high-performance computing system, we have additionally analyzed the absolute performance of our code.
In the following, we will concentrate on the single-node performance of our optimized code.

We show the performance analysis of the calculation sweeps on the single node kasper.
First, we use the roofline performance model to categorize our code in the memory- or compute-bound region~\cite{williams2009roofline}.
We use LIKWID\cite{likwid} to measure the maximum attainable bandwidth. On kasper we reach a bandwidth of approximately $\unit[35]{GiB/s}$, on one fh2 node we gain approximately $\unit[50]{GiB/s}$.
Since we are using a D3C7 stencil to swipe across the entire domain, four of the seven values to be loaded have already been loaded in the previous cell, so we can assume that only three values need to be loaded.
The remaining data values are already in the cache, see \S\ref{sec:cache} for details. The spatial data each holds the entire vector $\vec{u}$.
For the interpolation of the time-independent parameter data, $\unit[130]{Byte}$ are not located in the cache and have to be loaded for on lattice update.
The sweeps have to load $\unit[24]{Byte}$ per vector component.
Remember that we need three sweeps to process one time-step, so an average of $\unit[94.5]{Byte}$ for $P_3$ are loaded per lattice component update, $\unit[77.6]{Byte}$, $\unit[72.2]{Byte}$ for $P_7$, $P_{39}$, respectively.
If we only consider the memory bottleneck, we would get a theoretical peak-performance on fh2 of $\unit[50]{GiB/s}\cdot \unit[72.2]{Bytes/LCUP} = \unit[3\,785]{MLCUP/s}$ and $\unit[2\,527]{MLCUP/s}$ on kasper.
That is far away from what we measured---an indication that we are operating on the compute-bound side.
Counting the floating-point operations for one time-step, we get $392 + 40 v_\text{e} + 50 v_\text{o}\unit{FLOP}$, where $v_\text{e}$ is the number of even and $v_\text{o}$ the number of odd vector components.

So an average of $\unit[68.3]{FLOP}$ for $P_3$ are used per lattice component update, $\unit[50.5]{FLOP}$, $\unit[45.1]{FLOP}$ for $P_7$, $P_{39}$, respectively.
This results in an arithmetic intensity on the lower bound of $\unit[0.72]{FLOP/Byte}$ to $\unit[0.62]{FLOP/Byte}$ for $P_3$, $P_{39}$, respectively.
The Haswell CPU in kasper has an AVX base frequency of $\unit[2.7]{GHz}$~\cite{processor2017e5} and can perform $16$ floating-point operations with double precision per cycle.
This results in $\unit[43.2]{GFLOP/s}$ per core.
The achieved $\unit[139.1]{MLCUP/s}$ per core corresponds to $\unit[6.3]{GFLOP/s}$ and so to $\unit[15]{\%}$ peak-performance.

\subsubsection{Cache Effects}\label{sec:cache}
Even if the analysis in the previous section shows that our application is compute-bound, it is worth taking a look at the cache behavior.
Running large blocks will result in an excellent parallel scaling because of the computational time increase by $O(n^3)$ and the communication data only by $O(n^2)$.

\begin{figure}[tb!!]
  \centering
  \includegraphics[width=0.8\textwidth]{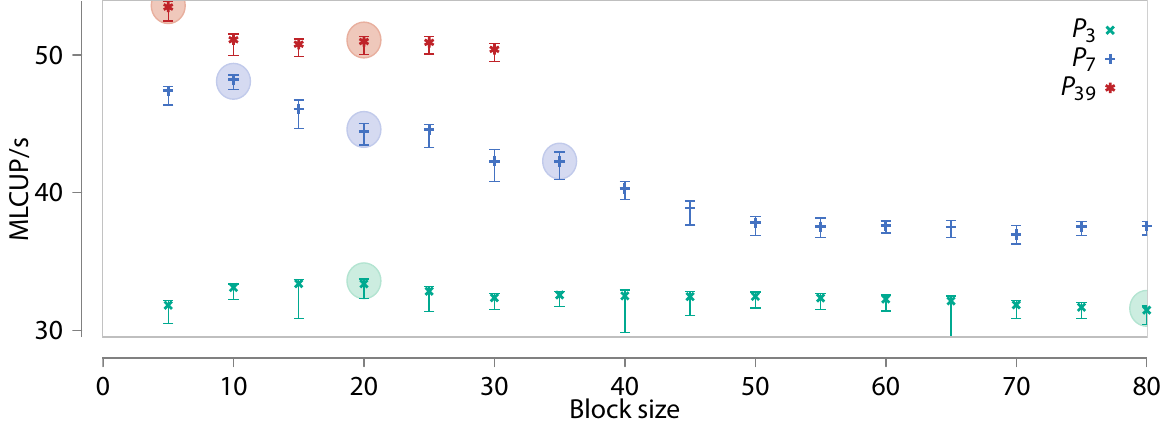}
  \caption{Performance of the calculation sweeps for different block sizes.}
  \label{fig:cache}
\end{figure}

To discover the cache behavior, we run 20 single jobs in parallel on one node on the fh2, this simulates the $\unit[2.5]{MiB}$ L3 cache per core.
Fig.~\ref{fig:cache} show the performance for different block sizes.
For $P_7$, a maximum block size of 13 fits into the L2 cache, here the largest performance can be seen. At a block size of 35, the performance drops, which can be explained by the fact that with a maximum block size of 40, three layers fit into the L3 cache.
For $P_3$, a block size of 20 still fits into the L2 cache, so here is the peak, up to a block size of 80 the performance remains almost constant after dropping firstly, this is the size where the three layers fit into the L3 cache.
The maximum for $P_{39}$ is at a block size of 5, here the three layers fit into the L3 cache.
We have not tested a smaller block size, because of the overhead of loops becomes too big.
We will use the marked block sizes for the scaling analysis in the following sections. The block size of 20 was chosen so that all three moment orders can be compared here.

\subsubsection{Scaling Results}
To examine the parallel scalability of our application, we consider weak scaling for different block sizes.
During one run, each process gets a block of the same size.
So we gain accurate scaling data that does not depend on any cache effects described in \S\ref{sec:cache}.
First, we look at one node of the fh2, and then at the performance across multiple nodes, with each node running 20 processes at the 20 cores.
We use up to 256 nodes, which are 5\,120 cores.
\begin{figure}[tb!]
  \centering
  \includegraphics[height=0.4\textwidth]{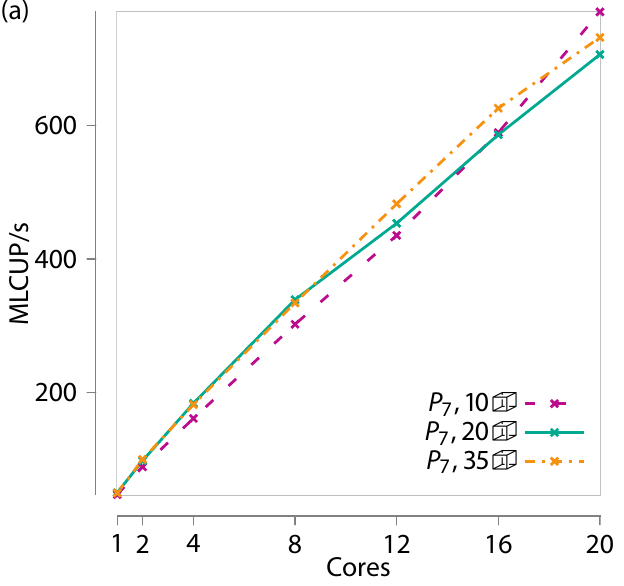}\hfill
  \includegraphics[height=0.4\textwidth]{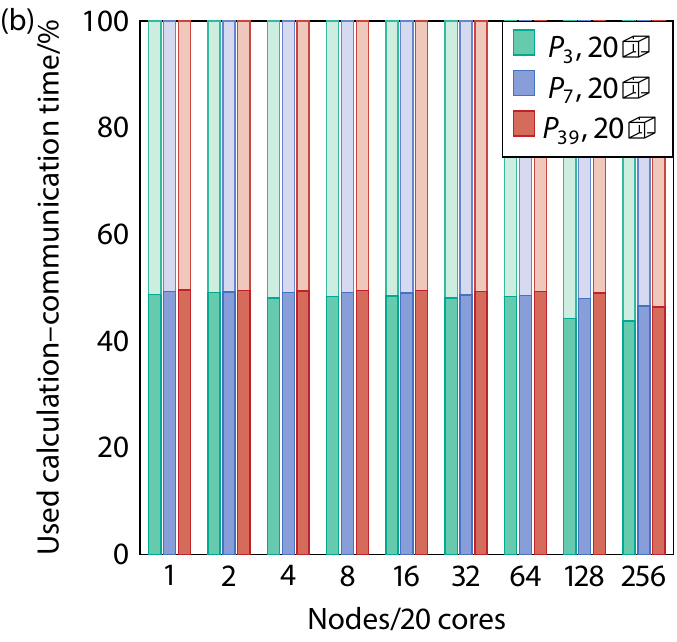}
  \caption{(a) Single Node scaling on fh2. (b) Calculation time (dark) versus communication time (light).}
  \label{fig:nodescale}
\end{figure}
Fig.~\ref{fig:nodescale}(a) shows $P_7$-runs with different block sizes, where the MPI processes distributed equally over the two sockets.
All three block sizes show similar, well-scaling behavior.
Moreover, the whole node does not reach the bandwidth limit of $\unit[3\,785]{MLCUP/s}$, which confirms that the application is on the compute-bound side.

Before conducting scaling experiments, we evaluate the various parts of the application.
Therefore, we show the amount of used calculation and communication time in Fig.~\ref{fig:nodescale}(b).
The calculation time for one time-step consists of the time used by two sweeps $S^\text{o}$ and one sweep $S^\text{e}$.
The communication time sums up the time used for the three halo exchanges.
A high communication effort of about $\unit[50]{\%}$ is necessary.
This proportion rarely changes for different vector lengths.

\begin{figure}[tb!]
  \centering
  \includegraphics[width=0.75\textwidth]{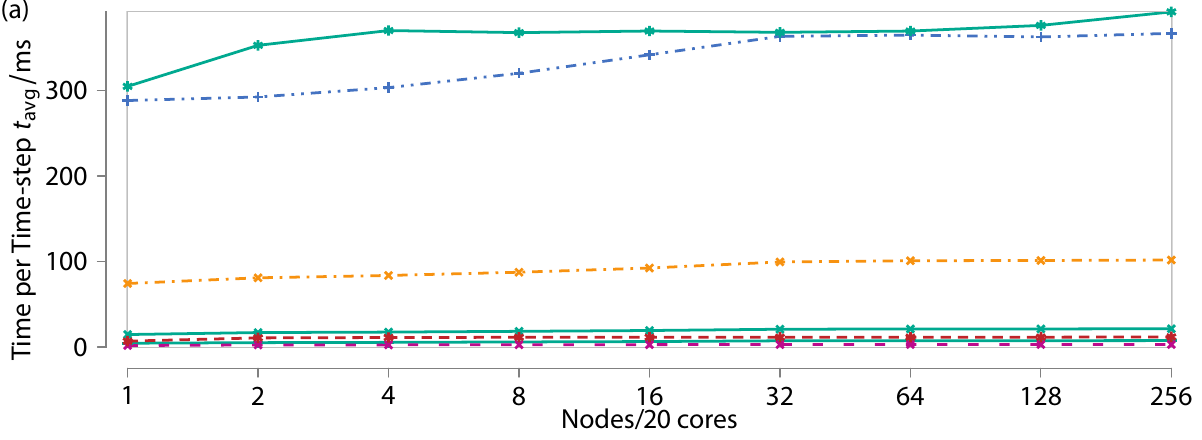}\\
  \includegraphics[width=0.75\textwidth]{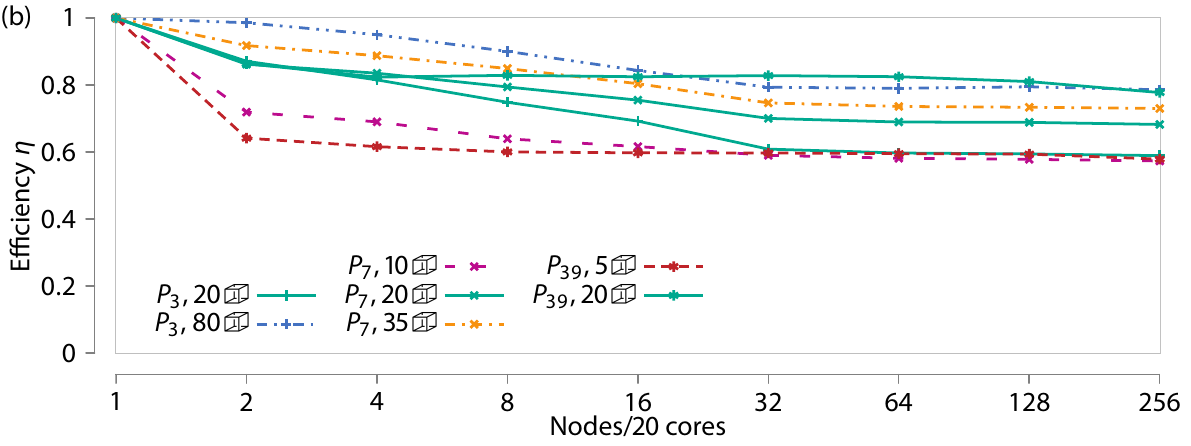}
  \caption{MPI scaling (a) average time per time-step and (b) efficiency on fh2 for up to 5\,120 cores.}
  \label{fig:scale}
\end{figure}
Fig.~\ref{fig:scale} shows the parallel scalability of the application for different vector lengths and block sizes.
The results of runs with one node are used as the basis for the efficiency calculations.
In (a) three regimes are identifiable, $P_3, 80$\mancube{} and $P_{39}, 20$\mancube{} are more expensive and take a long time.
$P_7, 35$\mancube{} is in the middle, and the remainder takes only a short average time per time-step.
As expected, this is also reflected in the efficiency in (b).
The expensive tasks scale slightly better with approximately $\unit[80]{\%}$ efficiency on 256 nodes, 5120 cores.
The shorter tasks still have approximately $\unit[60]{\%}$ efficiency.
From one to two nodes, there is a drop in some jobs; the required inter-node MPI communication can explain this.
From 32 nodes, the efficiency of all sizes is almost constant.
This is because a maximum of 23 nodes is connected to one switch, i.e., the jobs must communicate via an additional switch layer.
For runs on two to 16 nodes, the job scheduler can distribute the job to nodes connected to one switch but does not have to.

\subsection{Simulation Results}
\begin{figure}[tb!]
  \centering
  \includegraphics[height=3.74cm]{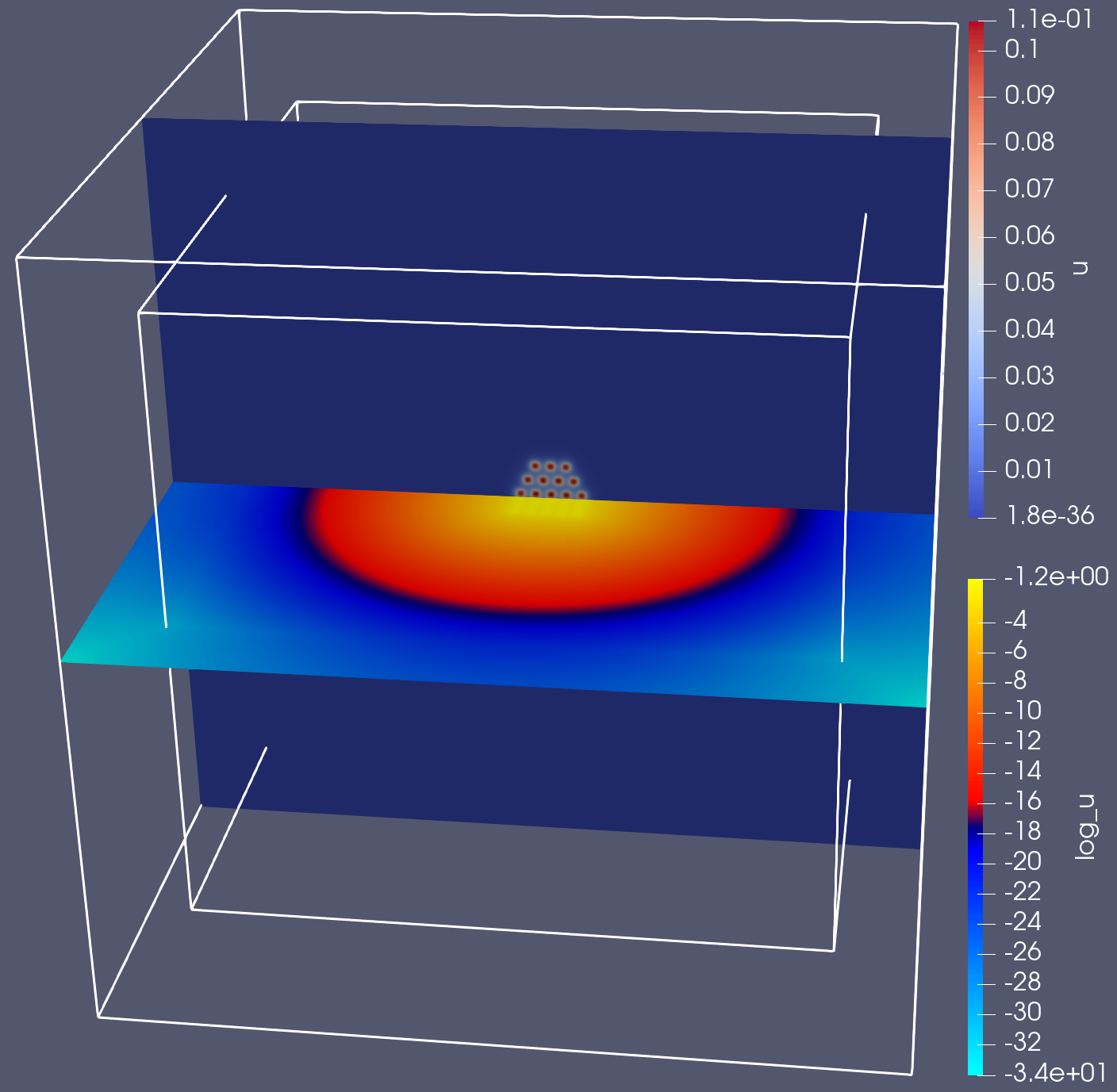} \hfill
  \includegraphics[height=3.74cm]{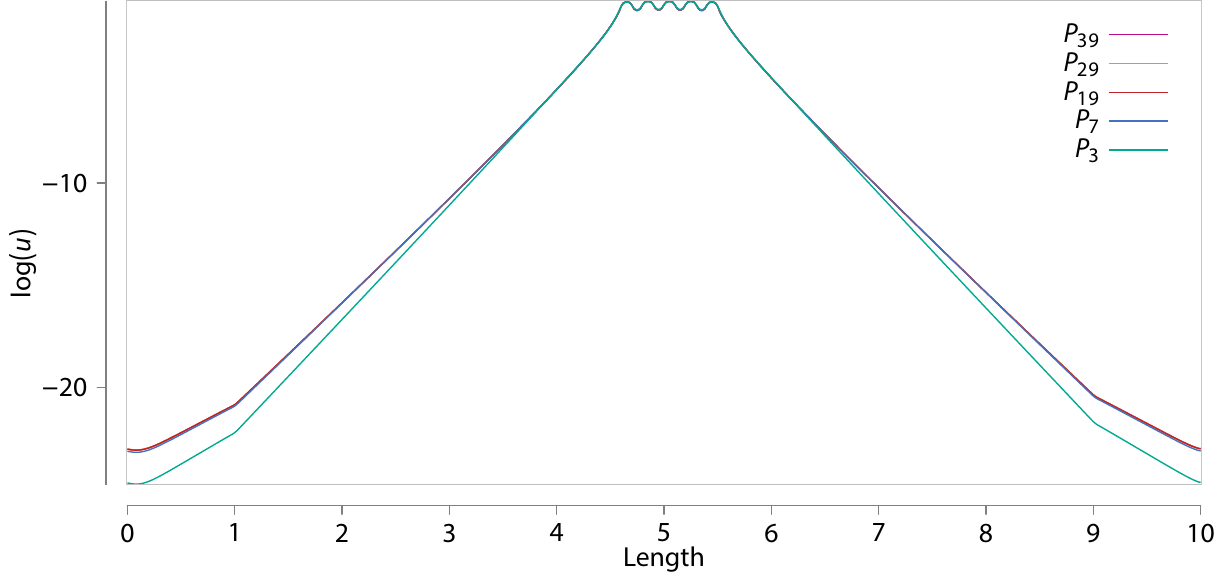}
  \caption{Left: Rod geometry surrounded by water and concrete.
  The vertical slice shows $u$ and the plane $\log_{10}(u)$.
  Right: Plot of the intensity $\log_{10}(u)$ over the section.}
  \label{fig:reaktor}
\end{figure}

With the parallelizability and scalability of the methodology and implementation established, we now showcase its applicability in a representative test example.
We consider a cube geometry that resembles radiation transport (albeit with simplified physics) in a nuclear reactor vessel, consisting of a reactor core with fuel rods, each $\unit[1]{cm}$ (5 grid-points) thick, surrounded by water (inner box in Fig.~\ref{fig:reaktor}, and concrete (outer box).
The non-dimensional material parameters are: source $q_0 = 2$, absorption $\Sigma_\text{a}^\text{w}=10, \Sigma_\text{a}^\text{c}=5$, scattering $\Sigma_\text{s}=1$.
The spatial resolution of the rod geometry and surrounding has a grid size of 500\mancube{}, which we compute on up to 2000 cores via moment resolutions $P_3$, $P_7$, $P_{19}$, $P_{29}$, and $P_{39}$, depicted in Fig.~\ref{fig:reaktor} right.
As one can see by comparing $P_N, N \geq 19$, the $P_{19}$ simulation is well-resolved.

\section{Conclusion}
We have developed and evaluated a massively parallel simulation code for radiation transport based on a moment model, which runs efficiently on current HPC systems.
With this code, we show that large domain sizes are now available.
Therefore, an HPC implementation is of crucial importance.
Starting from the reference implementation of \StaRMAP{} in \MATLAB{}, we have developed a new, highly optimized implementation that can efficiently run on modern HPC systems.
We have applied optimizations at various levels to the highly complex stencil code, including explicit SIMD vectorization.
Systematic performance engineering at the node-level resulted in a speedup factor of $4.35$ compared to the original code and $\unit[15]{\%}$ of peak performance at the node-level.
Besides, we have shown excellent scaling results for our code.

\section{Acknowledgments}
This work was performed on the supercomputer ForHLR funded by the Ministry of Science, Research and the Arts Baden-W{\"u}rttemberg and by the Federal Ministry of Education and Research.
B. Seibold wishes to acknowledge support by the National Science Foundation through grant DMS--1719640.

%
%
%
\bibliographystyle{splncs04}
\bibliography{bibliography,references_seibold}
%

\end{document}